\documentclass[aps,prd,preprint,superscriptaddress]{revtex4}
\usepackage{graphicx}
\usepackage{amsmath,amssymb}
\usepackage{bm}
\usepackage[caption=false]{subfig}
\usepackage{color}

\def\p{\partial}

\def\=:{=\hspace{-.7em}\raisebox{1.1ex}{.}\hspace{.1em}\raisebox{-0.2ex}{.} }

\newcommand {\beq}{\begin{eqnarray}}
\newcommand {\eeq}{\end{eqnarray}}
\newcommand {\non}{\nonumber\\}

\begin{document}

\preprint{NORDITA-2014-144}

\title{D-brane solitons in various dimensions}

\author{Sven Bjarke Gudnason}
\affiliation{Nordita, KTH Royal Institute of Technology and Stockholm
  University, Roslagstullsbacken 23, SE-106 91 Stockholm, Sweden
}
\author{Muneto Nitta}
\affiliation{Department of Physics, and Research and Education Center
  for Natural Sciences, Keio University, Hiyoshi 4-1-1, Yokohama,
  Kanagawa 223-8521, Japan 
}

\date{\today}
\begin{abstract}
We construct a D-brane soliton, a composite topological soliton
sharing some properties with a D-brane, in a Skyrme model in 4+1
dimensions, in which Skyrmions are strings ending on a domain wall.
We further generalize this D-brane soliton to diverse dimensions.
A string, carrying the $\pi_{N-1}$ topological charge, ends on a domain wall in an O$(N)$ model with higher-derivative terms in $N+1$
dimensions. 
\end{abstract}


\maketitle

\section{Introduction}

Dirichlet (D-)branes are solitonic objects in string theory, 
on which fundamental strings can end \cite{Polchinski:1995mt}.
Since their discovery, D-branes 
have provided a fundamental tool to study 
non-perturbative dynamics of string theory and 
even quantum field theory such as 
a brane realization \cite{Witten:1997ep} 
and via the AdS/CFT correspondence 
\cite{Maldacena:1997re}.
The Dirac-Born-Infeld (DBI) action describes 
the collective-coordinate motion of D-branes.
In the DBI action, 
endpoints of fundamental strings 
can be regarded as solitonic excitations called 
BIons \cite{Callan:1997kz}. 
After some attempts in field theory 
to mimic D-branes \cite{Carroll:1997pz},
constructing 
the first exact solution of a composite soliton looking like a D-brane  
was achieved in Ref.~\cite{Gauntlett:2000de},
in which a lump string ends on a domain wall 
\cite{Abraham:1992vb} 
in a supersymmetric ${\mathbb C}P^1$ 
nonlinear sigma model. 
When one looks at this solution from the domain wall effective theory,
it reproduces a BIon, and therefore this solution was named 
a D-brane soliton. 
The D-brane soliton was 
promoted to that in supersymmetric QED, 
which is a U(1) gauge theory  
coupled with two charged scalar fields 
\cite{Shifman:2002jm,Shifman:2007ce}. 
More general D-brane solitons 
were constructed in
supersymmetric ${\mathbb C}P^n$ and Grassmann sigma models, 
and corresponding 
supersymmetric U$(N)$ gauge theories \cite{Isozumi:2004vg},
in which exact solutions with an arbitrary number of strings 
at an arbitrary position  
stretched between multiple domain walls  
\cite{Gauntlett:2000ib,Isozumi:2004jc}
were found 
(for a review see Ref.~\cite{Eto:2006pg}).
Low-energy dynamics, such as scattering of 
strings stretched between branes, 
was studied in the moduli-space approximation 
\cite{Eto:2008mf}. 
In Ref.~\cite{Isozumi:2004vg},
a negative monopole charge was  
found at the endpoint of a string,
which was later named a boojum \cite{Sakai:2005sp}; see also
Ref.~\cite{Auzzi:2005yw},
and the boojum charge was also reproduced in 
a domain-wall effective action \cite{Eto:2006uw}.
Strings stretched between 
a brane and anti-brane pair, 
their approximate solutions,  
and a fate after a pair annihilation were discussed 
in Ref.~\cite{Nitta:2012kk}.
A wall-vortex junction in the large magnetic flux limit was studied in 
Ref.~\cite{Bolognesi:2005zr}.
 
The term boojum was taken from 
condensed matter physics, 
and in fact, boojums have been already studied  
in various condensed matter systems
\cite{Volovik:2003} such as
nematic liquids \cite{Volovik:1983}, 
superfluids at the edge of a container filled with $^4$He, 
at the A-B phase boundary of $^3$He \cite{Blaauwgeers:2002}, 
multi-component Bose-Einstein condensates (BEC) 
of ultracold atomic gases \cite{Takeuchi:2006}, 
spinor BECs \cite{Borgh:2012es},
and in even dense quark matter \cite{Cipriani:2012hr}.
Among others, in particular, 
D-brane solitons accompanied by boojums of 
the same type were constructed
in two-component BECs
\cite{Kasamatsu:2010aq}. 

Strings in D-brane solitons found thus far 
are of codimension two. 
In this paper, we offer a very simple model 
admitting strings (of higher codimensions) 
ending on a domain wall in higher dimensions. 
It is an O$(N)$ nonlinear sigma model 
with higher-derivative (Skyrme-like) term(s)
in $N+1$ dimensions
and a quadratic potential term with two vacua and thus admitting a
domain wall. 
The O(3) model is a baby-Skyrme model \cite{Piette:1994ug} 
with a quadratic potential 
\cite{Weidig:1998ii}
in 3+1 dimensions, while 
the O(4) model is the Skyrme model 
with the quadratic potential \cite{Kudryavtsev:1999zm,Nitta:2012wi,
Gudnason:2013qba,Nitta:2012rq,Gudnason:2014hsa}
in 4+1 dimensions.
The model admits a baby-Skyrmion string with 
$\pi_2$ lump charge ($N=3$),
a Skyrmion string with $\pi_3$ Skyrmion (baryon) charge ($N=4$), 
or higher dimensional analogs with 
$\pi_{N-1}$ topological charge. 
We numerically construct solutions 
of these D-brane solitons for $N=3,4,5,6$.
For the O(3) model in 3+1 dimensions, 
we construct a baby-Skyrmion string ending on a domain wall, 
which is baby-Skyrmion version of the prototype 
of a lump-string ending on a wall
\cite{Gauntlett:2000de}.
For the O(4) model in 4+1 dimensions, 
we construct a Skyrmion string 
with $\pi_{3}$ topological (baryon) charge, 
ending on a domain wall.
For the O($N$) model in $N+1$ dimensions, 
we have a higher dimensional Skyrmion-like string of codimensions $N-1$,  
supported by the $\pi_{N-1}$ topological charge, 
ending on a domain wall. For $N>3$ higher-derivative terms are needed
to prevent the string from collapsing to a singular solution. 
We study the shapes of domain walls pulled by such (finite) strings. 
It is known that the shape is logarithmic for the O(3) model without 
the fourth-order derivative term. 
We find that once the higher-derivative terms are considered, the
shape is $1/\rho^\#$, where the power $\#$ is fitted to be about 5 and
perhaps is universal. 

This paper is organized as follows.
Sec.~\ref{sec:model} presents our model: 
an O$(N)$ sigma model with higher-derivative terms.
In Sec.~\ref{sec:sigmamodel}, 
we review the D-brane soliton in the O(3) model 
without any higher-derivative terms in $d=3+1$.
In Sec.~\ref{sec:skyrmemodel}
we construct the D-brane soliton in 
the Skyrme model for O(3) in $d=3+1$ dimensions and for O(4) in
$d=4+1$ dimensions.
In Sec.~\ref{sec:6thmodel}
we generalize the construction to $5+1$ and $6+1$ dimensions,
necessitating an even higher-order derivative term; explicitly we
consider a sixth-order term for $N=5,6$ in $d=5+1$ and $6+1$,
respectively. 
Section \ref{sec:summary} is devoted to a summary and discussion.

\newpage

\section{The model}\label{sec:model}

We consider the O$(N)$ sigma model with higher-derivative terms in
$N+1$ dimensions whose Lagrangian density reads
\beq
\mathcal{L} = 
-\frac{1}{2}\p_\mu \mathbf{n} \cdot \p^\mu\mathbf{n} 
+ c_4 \mathcal{L}_4
+ c_6 \mathcal{L}_6
- V,
\label{eq:Lmother}
\eeq
where $\mu=1,\ldots,N$; $\mathbf{n}=(n_1,\ldots,n_N)^{\rm T}$;
$\mathbf{n}\cdot\mathbf{n}=1$ and 
\begin{align}
\mathcal{L}_4 &= 
-\frac{1}{4}\left(\p_\mu \mathbf{n} \cdot \p^\mu\mathbf{n}\right)^2
+\frac{1}{4}\left(\p_\mu \mathbf{n} \cdot \p_\nu\mathbf{n}\right)^2,\\
\mathcal{L}_6 &= 
-\left(\p_\mu n^{[a} \p_\nu n^{b} \p_\rho n^{c]}\right)^2 
\label{eq:L6}\\
&=
-\frac{1}{6}\left(\p_\mu \mathbf{n} \cdot \p^\mu\mathbf{n}\right)^3
+\frac{1}{2}\left(\p_\mu \mathbf{n} \cdot \p^\mu\mathbf{n}\right)
  \left(\p_\nu \mathbf{n} \cdot \p_\rho\mathbf{n}\right)^2
-\frac{1}{3}\left(\p_\mu \mathbf{n} \cdot \p^\nu\mathbf{n}\right)
  \left(\p_\nu \mathbf{n} \cdot \p^\rho\mathbf{n}\right)
  \left(\p_\rho \mathbf{n} \cdot \p^\mu\mathbf{n}\right), \nonumber
\end{align}
where $a,b,c=1,\ldots,N$ and the antisymmetrization is defined as
\beq
T^{[abc]} \equiv \frac{1}{3!}\left(
T^{abc} + T^{bca} + T^{cab} - T^{cba} - T^{bac} - T^{acb}\right).
\eeq
We are using the mostly-positive metric.
In the absence of the potential, the symmetry is 
O($N$) which is spontaneously broken to O($N-1$).
The target space of the sigma model is 
\beq 
  {\rm O}(N)/{\rm O}(N-1) \simeq S^{N-1}.
\eeq 
We consider the potential,  given by 
\beq
V = \frac{1}{2} m^2 (1-n_N^2).
\label{eq:potential}
\eeq
The vacua are 
\begin{align}
+ \quad : \qquad & n_N = +1, \label{eq:vac+}\\
- \quad : \qquad & n_N = -1. \label{eq:vac-}
\end{align}
The above potential breaks the O$(N)$ symmetry to O$(N-1)$ explicitly.

Note that the Lagrangian density \eqref{eq:L6} is the baryon density
squared when $N=4$ and is the basis of the BPS Skyrme model
\cite{Adam:2010fg}. In the present formulation of the term, $N$ can be
larger than $4$ but then the term no longer represents the $\pi_{N-1}$
charge. 

We will consider a domain wall extended in the $z\equiv x^N$
direction, which interpolates between the $-$ and the $+$ vacua of
Eqs.~\eqref{eq:vac+} and \eqref{eq:vac-}. 
The domain-wall solution is given by
\beq
n_N = \tanh(m z), 
\eeq 
and is an exact solution: 
it is the sine-Gordon soliton
(for the O(3) model \cite{Abraham:1992vb}, 
the O(4) model 
\cite{Kudryavtsev:1999zm,Losev:2000mm,Nitta:2012wi,Gudnason:2013qba,Gudnason:2014hsa},
and the O($N$) model \cite{Nitta:2012rq}).

In this paper we are interested in the soliton junction composed by
the latter domain wall and a ``string'' carrying $\pi_{N-1}$ charge. 
This is possible because zero modes (moduli) are localized on the
domain wall, which originate from the spontaneously broken O($N-1$)
symmetry in the presence of the wall. 
Those moduli are U(1) for the O(3) model \cite{Abraham:1992vb}, 
$S^2$ for the O(4) model 
\cite{Losev:2000mm,Nitta:2012wi,Gudnason:2013qba,Gudnason:2014hsa},
and $S^{N-2}$ for the O($N$) model \cite{Nitta:2012rq}.
One could construct textures (Skyrmions) 
supported by $\pi_{N-2} (S^{N-2})\simeq {\mathbb Z}$
\cite{Auzzi:2006ju,Nitta:2012xq,Nitta:2012wi,Gudnason:2013qba,Gudnason:2014hsa,Nitta:2012rq,Garaud:2012,Jennings:2013aea}, 
which are localized on the wall. 
Instead, here, we discuss defects again
supported by $\pi_{N-2} (S^{N-2})\simeq {\mathbb Z}$. 
As we will see, these defects actually extend in 
the direction perpendicular to the wall, 
and it turns out that they are Skyrmion strings 
supported by the $\pi_{N-1}(S^{N-1})\simeq {\mathbb Z}$ 
in the bulk.

An appropriate Ansatz for the configuration is
\beq
\mathbf{n} = \left(\frac{\tilde{\mathbf{x}}}{\rho}\sin f(\rho,z), 
  \cos f(\rho,z)\right)^{\rm T},
\eeq
where $\tilde{\mathbf{x}}=(x^1,\ldots,x^{N-1})$;
$\rho^2=(x^1)^2+\cdots+(x^{N-1})^2$ is the $(N-1)$-dimensional radial 
coordinate and $z\equiv x^N$. Inserting the above Ansatz into the
Lagrangian density \eqref{eq:Lmother} we obtain the following static
Lagrangian density
\begin{align}
-\mathcal{L} &= 
\frac{1}{2}f_\rho^2
+\frac{1}{2}f_z^2
+\frac{N-2}{2\rho^2}\sin^2 f
+c_4\left[
  \frac{N-2}{2\rho^2}\sin^2 f\left(f_\rho^2 + f_z^2\right)
  +\frac{(N-3)(N-2)}{4\rho^4}\sin^4 f
\right] \non
&\phantom{=\ }
+c_6\left[
  \frac{(N-3)(N-2)}{2\rho^4}\sin^4 f\left(f_\rho^2 + f_z^2\right)
  +\frac{(N-4)(N-3)(N-2)}{6\rho^6}\sin^6 f
\right] \non
&\phantom{=\ }
+\frac{1}{2}m^2\sin^2 f,
\label{eq:Lfmother}
\end{align}
where $f_x\equiv \p_x f$ and the equation of motion reads
\begin{align}
f_{\rho\rho}
+f_{zz}
+\frac{N-2}{\rho} f_\rho
+c_4\frac{N-2}{\rho^2}\sin^2 f\left(
  f_{\rho\rho} + f_{zz} + \frac{N-4}{\rho} f_\rho\right)
+c_4\frac{N-2}{2\rho^2}\sin(2f)\left(f_\rho^2 + f_z^2\right) \non
\mathop+c_6\frac{(N-3)(N-2)}{\rho^4}\sin^4 f\left(
  f_{\rho\rho} + f_{zz} + \frac{N-6}{\rho} f_\rho\right) \non
\mathop+c_6\frac{(N-3)(N-2)}{\rho^4}\sin^2f\sin(2f)\left(
  f_\rho^2 + f_z^2\right)
-\frac{N-2}{2\rho^2}\sin(2f) \non
\mathop-c_4\frac{(N-3)(N-2)}{2\rho^4}\sin^2 f\sin(2f)
-c_6\frac{(N-4)(N-3)(N-2)}{2\rho^6}\sin^4 f\sin(2f) \non
\mathop-\frac{1}{2}m^2 \sin 2f = 0.
\label{eq:f_eom}
\end{align}

The $\pi_{N-1}$ charge is given by
\begin{align}
C &= \frac{\Gamma\left(\frac{N}{2}\right)}{2\pi^{\frac{N}{2}}} 
\int d^{N-1} x \;\frac{1}{(N-1)!}
\epsilon^{i_1\cdots i_{N-1}}\epsilon^{a_1\cdots a_N}
\p_{i_1}n_{a_1} \cdots \p_{i_{N-1}} n_{a_{N-1}} n_{a_N} \non
&= \frac{\Gamma\left(\frac{N}{2}\right)}
{\sqrt{\pi}\Gamma\left(\frac{N-1}{2}\right)}
\int d\rho \; \sin^{N-2}(f) \p_\rho f \non
&= \left.-\cos f \left(\mathop{\rm sign}\sin f\right)^{N-1} 
\frac{\Gamma\left(\frac{N}{2}\right)}
  {\sqrt{\pi}\Gamma\left(\frac{N-1}{2}\right)} \;
{}_2F_1\left(\frac{1}{2},\frac{3-N}{2};\frac{3}{2};\cos^2 f\right)
\right|_{f(0)}^{f(\infty)} = -1,
\label{eq:stringcharge}
\end{align}
where $i_{\#}=1,\ldots,N-1$; $a_{\#}=1,\ldots,N$; $\Gamma$ is the
gamma function and ${}_2F_1$ is the usual hypergeometric function.
(Note that the overall sign is chosen for convenience and is opposite
to the conventional choice). 
With the boundary conditions $f(0)=0$ and $f(\infty)=\pi$, the above
charge integrates to $C=-1$ for all $N\geq 3$, as shown in the last
step of the last line. 
The charge density is given by
\beq
\mathcal{C} = \frac{1}{\rho^{N-2}} \sin^{N-2}(f) \p_\rho f.
\eeq

The reason for including higher-derivative terms in the action
\eqref{eq:Lmother} is to prevent the ``string'' from collapsing to a
singular solution (at a finite distance from the domain wall). 
Let us consider a scaling argument for just the coordinates transverse
to the domain wall, $\tilde{\mathbf{x}}\to\mu\tilde{\mathbf{x}}$.
The energy thus scales as
\begin{align}
E &\to -\int d^{N-1} x\; \Big[
\mu^{3-N} \mathcal{L}_2^\rho
+\mu^{1-N} \mathcal{L}_2^z \non
&\phantom{= -\int d^{N-1} x\; \Big[\ }
+\mu^{5-N} c_4 \mathcal{L}_4^\rho
+\mu^{3-N} c_4 \mathcal{L}_4^z
+\mu^{7-N} c_6 \mathcal{L}_6^\rho
+\mu^{5-N} c_6 \mathcal{L}_6^z
-\mu^{1-N} V
\Big],
\label{eq:scaling_argument}
\end{align}
where $\mathcal{L}_d^{\rho,z}$ denotes the part of the Lagrangian
density of $d$th order in derivatives having two derivatives in $\rho$
and $z$, respectively \footnote{The remaining derivatives are angular
  derivatives which in the Ansatz with $(N-1)$-dimensional spherical
  symmetry reduces to power of the $(N-1)$-dimensional radius,
  $\rho$. }.
Stability of the solitonic solution requires a positive power of $\mu$
for at least one term. If the power is zero (i.e.~$\mu^0$), the term
is classically conformal and cannot provide stabilization. 

For $N=3$ we can have a finite-size (lump-charged) string with just
the Dirichlet term and thus no higher-derivative terms ($c_4=c_6=0$). 
The apparent instability in the $z$ coordinate does not affect the
solution because of factorization (see the next section). Note also
that the $\rho$ part of the energy is classically conformal; a
characteristic of lumps. 
For $N=4$ and just the Skyrme term turned on ($c_4=1$ and $c_6=0$)
we have a stable finite-size Skyrmion-charged string.
At each constant-$z$ slice, 
a domain wall becomes a spherical domain wall 
studied in Ref.~\cite{Gudnason:2013qba} 
which is nothing but a Skyrmion. 
For $N=5,6$ we need the sixth-order derivative term for stabilizing
the string. 
We will consider these three cases in turn in the next sections.

\section{The pure sigma model}\label{sec:sigmamodel}

In this section, we use only the kinetic term and the potential,
i.e.~the Lagrangian density \eqref{eq:Lfmother} with $c_4=c_6=0$. 

This system is special for $N=3$ where it corresponds to an integrable 
sector in the supersymmetric O$(3)$ sigma model
\cite{Gauntlett:2000ib}. 
For $N=2$, the system is again integrable, but is somehow trivial as
it describes two domain walls orthogonally assembled. 

In order to uncover the domain-wall structure of the system, let us
change variables as 
\beq
f = 2\arctan g,
\eeq
which gives us the non-linear equation of motion (we keep $N$ explicit
for illustrative purposes here)
\begin{align}
(1+g^2) g_{\rho\rho}
- 2g g_\rho^2
+ \frac{N-2}{\rho}(1+g^2)g_\rho
+ (1+g^2) g_{zz}
- 2g g_z^2
- \frac{N-2}{2\rho^2} g(1-g^2) \non
- m^2 g(1-g^2) = 0.
\end{align}
Regrouping this, we get
\begin{align}
g_{\rho\rho}
+\frac{N-2}{\rho} g_\rho
+g_{zz}
-\frac{N-2}{\rho} g
-m^2 g = 0,
\label{eq:linear_o3_dbrane}
\end{align}
with the nonlinear constraint (assuming the above equation is
satisfied) 
\beq
\frac{g_\rho^2}{g}
-\frac{N-2}{\rho^2} g
+\frac{g_z^2}{g}
-m^2 g = 0,
\eeq
and inserting a factorizing Ansatz $g=R(\rho)Z(z)$ into the latter
constraint, we get 
\beq
R = \rho^{\pm\sqrt{N-2}}, \qquad
Z = e^{\pm m (z-z_0)}.
\eeq
Inserting this into Eq.~\eqref{eq:linear_o3_dbrane} yields ($N\geq 2$)
\beq
\mp\sqrt{N-2}
\pm (N-2)^{\frac{3}{2}} = 0, 
\eeq
which determines $N=3$ (or the trivial solution $N=2$ which is
physically not so interesting). 
The exact solution in the nonlinear sigma model case for $N=3$ thus
reads 
\beq
g = r^{\pm 1} \exp\{\pm m (z - z_0)\}.
\eeq
The two signs are independent of each other and all four possibilities
are solutions to the equation of motion. They are however not all
physically different as the Lagrangian is invariant under $f\to\pi-f$
which corresponds to $g\to g^{-1}$. Using this fact, we see that there
are two distinct configurations which we can think of as a wall
junction and anti wall junction. These two are related by sending the 
coordinate $z\to -z$.

Factorization is possible when $N=3$ as we have just shown
above, but only possible when $N=3$. For illustrative purposes, let us
implement the domain-wall structure explicitly by setting 
\beq
g = \exp\{m z\} h(\rho,z),
\eeq
and study the string solutions on both sides of the domain wall. 
Notice that $m\to\pm m$ and $z\to z-z_0$ recovers the domain
wall/anti domain wall and position modulus, respectively. In order not 
to clutter the notation too much in the following, we will just use
$e^{mz}$. 
The equation of motion can now be written as
\begin{align}
h_{\rho\rho}\left(1 + e^{2m z} h^2\right)
- 2e^{2m z} h h_\rho^2 
+ \frac{N-2}{\rho}\left(1 + e^{2m z}h^2\right) h_\rho \non
\mathop+ h_{zz}\left(1 + e^{2m z} h^2\right)
- 2e^{2m z} h h_z^2
+ 2m\left(1 - e^{2m z} h^2\right) h_z \non
\mathop- \frac{N-2}{\rho^2}\left(1 - e^{2m z}h^2\right) h = 0.
\label{eq:eom_h}
\end{align}
The field $h$ will describe the junction in the (fixed) background of
the domain wall which is generally a solution to the above PDE and
hence a function of both $\rho$ and $z$. 
Taking the limit $z\to\infty$, the equation of motion \eqref{eq:eom_h}
becomes independent of $z$: 
\beq
h_{\rho\rho}
- \frac{2h_\rho^2}{h}
+ \frac{N-2}{\rho} h_\rho 
+ \frac{N-2}{\rho^2} h = 0,
\eeq
and a power function Ansatz $h=\rho^b$ yields the following two
solutions 
\beq
b_+ = N-2, \qquad  
b_- = -1.
\eeq
Taking now the limit $z\to-\infty$, the equation of motion
\eqref{eq:eom_h} becomes again independent of $z$:
\beq
h_{\rho\rho}
+ \frac{N-2}{\rho} h_\rho 
- \frac{N-2}{\rho^2} h = 0,
\eeq
and the power function Ansatz $h=\rho^b$ now yields the two solutions 
\beq
b_+ = 1, \qquad
b_- = -N+2.
\eeq
Having two different signs on each side of the domain wall corresponds
to a composite soliton made of a wall and an anti wall and thus is not
a solution on the fixed background. Therefore we need to pick the same
sign on each side of the domain wall, which corresponds to choosing a
string or an anti string (or alternatively which direction the string
is pointed).
The factorization is again visible for $N=3$ because the power
function Ansatz is the same on both sides of the domain wall (and in
fact as we showed earlier, it is a solution in all space). 
In principle we could contemplate a solution interpolating the two
different power functions when $N\geq 4$, but as shown by a scaling
argument in Sec.~\ref{sec:model} (see
Eq.~\eqref{eq:scaling_argument}), such solution will have a singular
(i.e.~vanishing thickness) string and the junction will also be point
like. We can blow up such solutions by adding higher-derivative terms,
as shown in Eq.~\eqref{eq:scaling_argument}. This will be the topic of
the next sections.

\section{The Skyrme model}\label{sec:skyrmemodel}

In this section, we turn on the Skyrme term in the Lagrangian density
\eqref{eq:Lfmother}, viz.~$c_4=1$ and $c_6=0$. This will allow for
stable finite-size strings for $N=4$ as shown by the scaling argument
in Eq.~\eqref{eq:scaling_argument}. 
The equation of motion \eqref{eq:f_eom}, in this case, is not
integrable and we need to turn to numerical methods to obtain
solutions. 

We will employ a finite-difference scheme on a quadratic square
lattice with $256^2$ lattice sites and relax initial guesses with the
relaxation method. 

For completeness, we also calculate the case of $N=3$ with the Skyrme
term, which makes the string thicker than in the sigma model case. 

\begin{figure}[!thp]
\begin{center}
\captionsetup[subfloat]{labelformat=empty}
\mbox{
\subfloat[$n_3=\cos f$]{\includegraphics[width=0.33\linewidth]{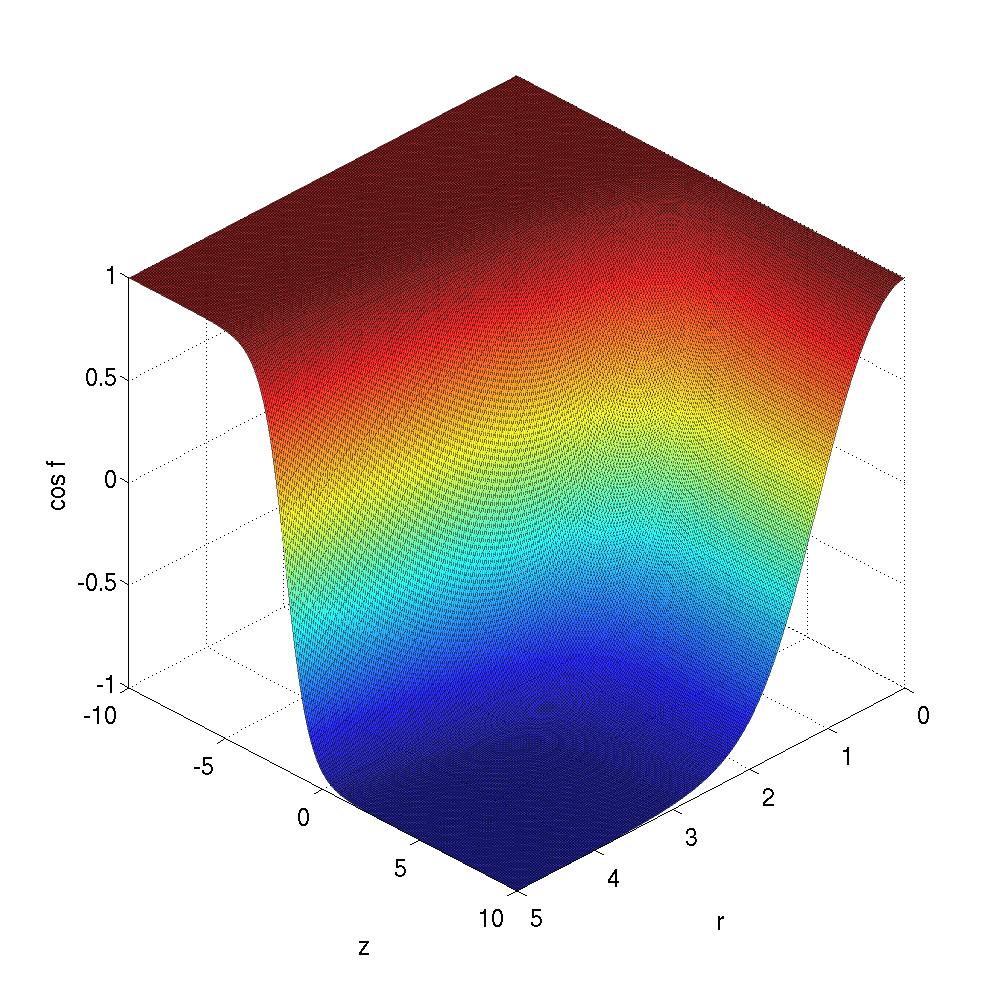}}
\subfloat[$\log(1+\mathcal{E})$]{\includegraphics[width=0.33\linewidth]{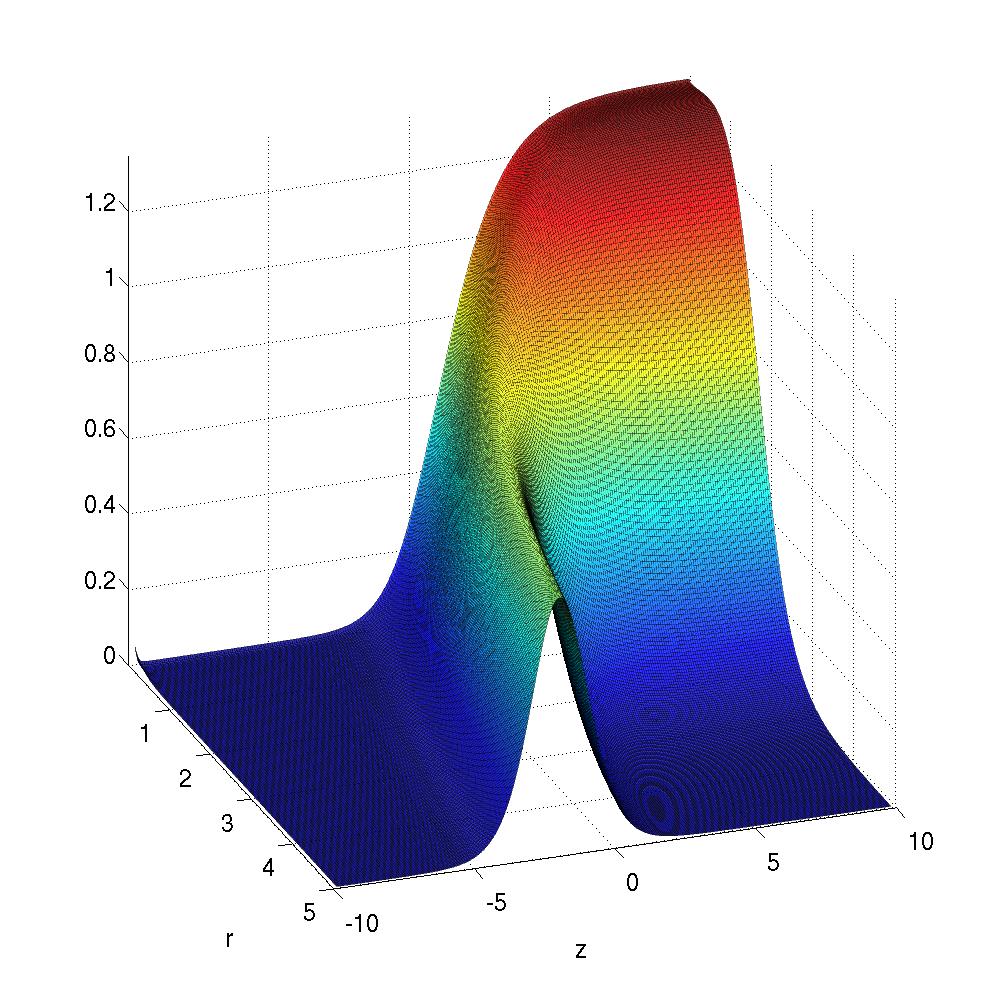}}
\subfloat[$\pi_2$ charge]{\includegraphics[width=0.33\linewidth]{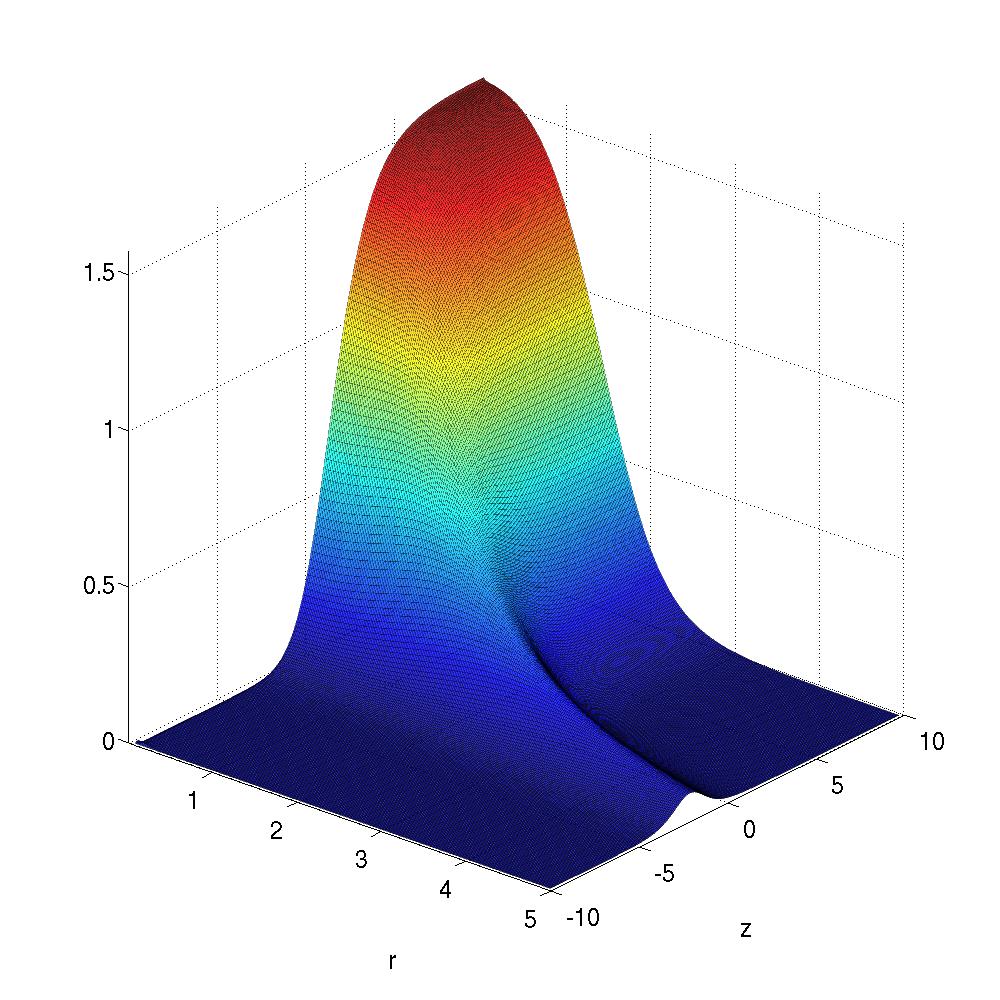}}}
\caption{${\rm O}(3)$ soliton junction of a domain wall (in the $z$
  direction) and a lump string carrying $\pi_2$ charge. The three
  panels show the field $n_3$, the energy density on a logarithmic
  scale and the charge density. }
\label{fig:o3}
\end{center}
\end{figure}

\begin{figure}[!thp]
\begin{center}
\captionsetup[subfloat]{labelformat=empty}
\mbox{
\subfloat[$n_4=\cos f$]{\includegraphics[width=0.33\linewidth]{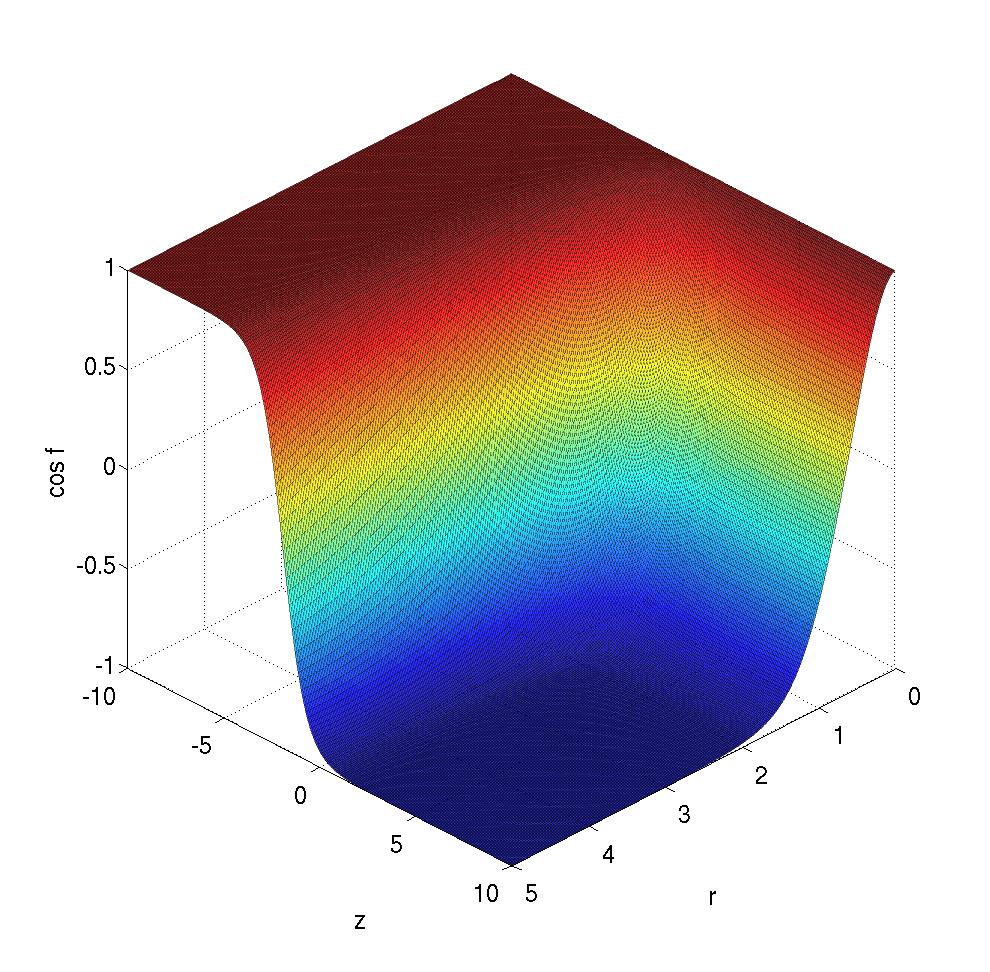}}
\subfloat[$\log(1+\mathcal{E})$]{\includegraphics[width=0.33\linewidth]{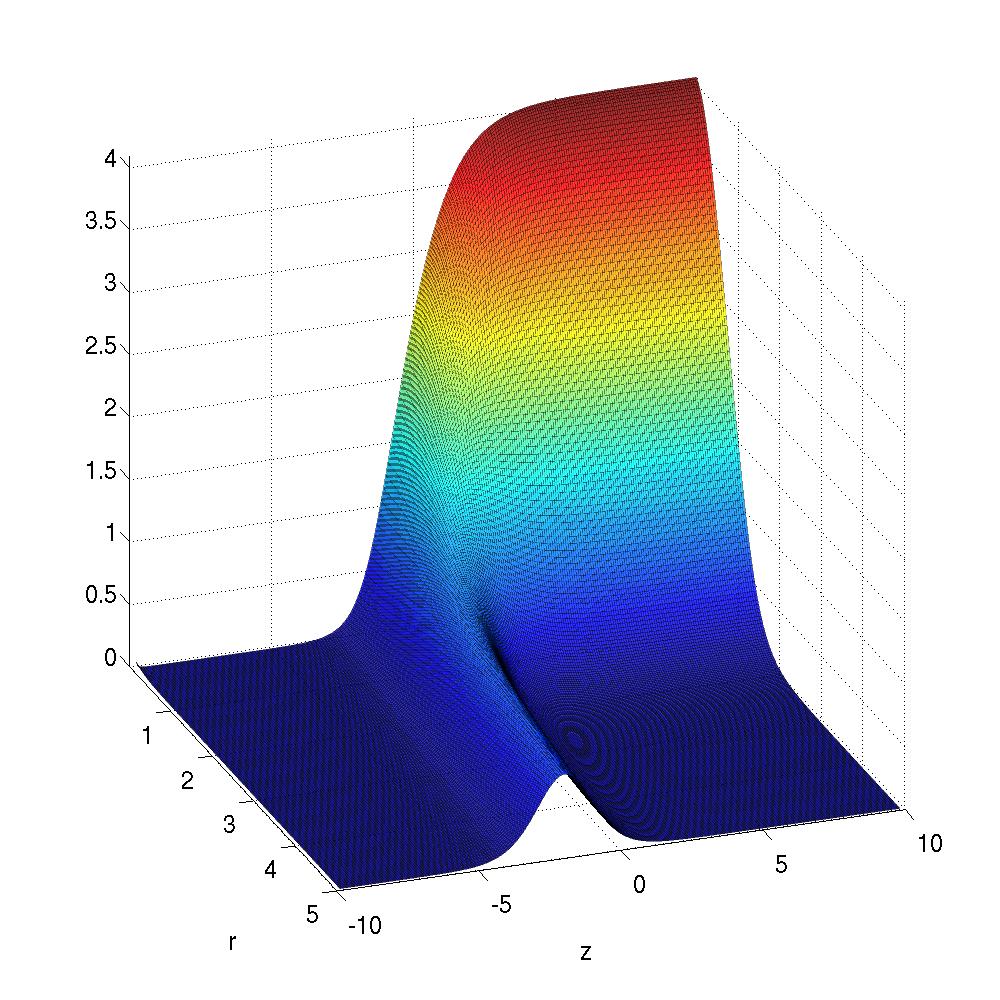}}
\subfloat[$\pi_3$ charge]{\includegraphics[width=0.33\linewidth]{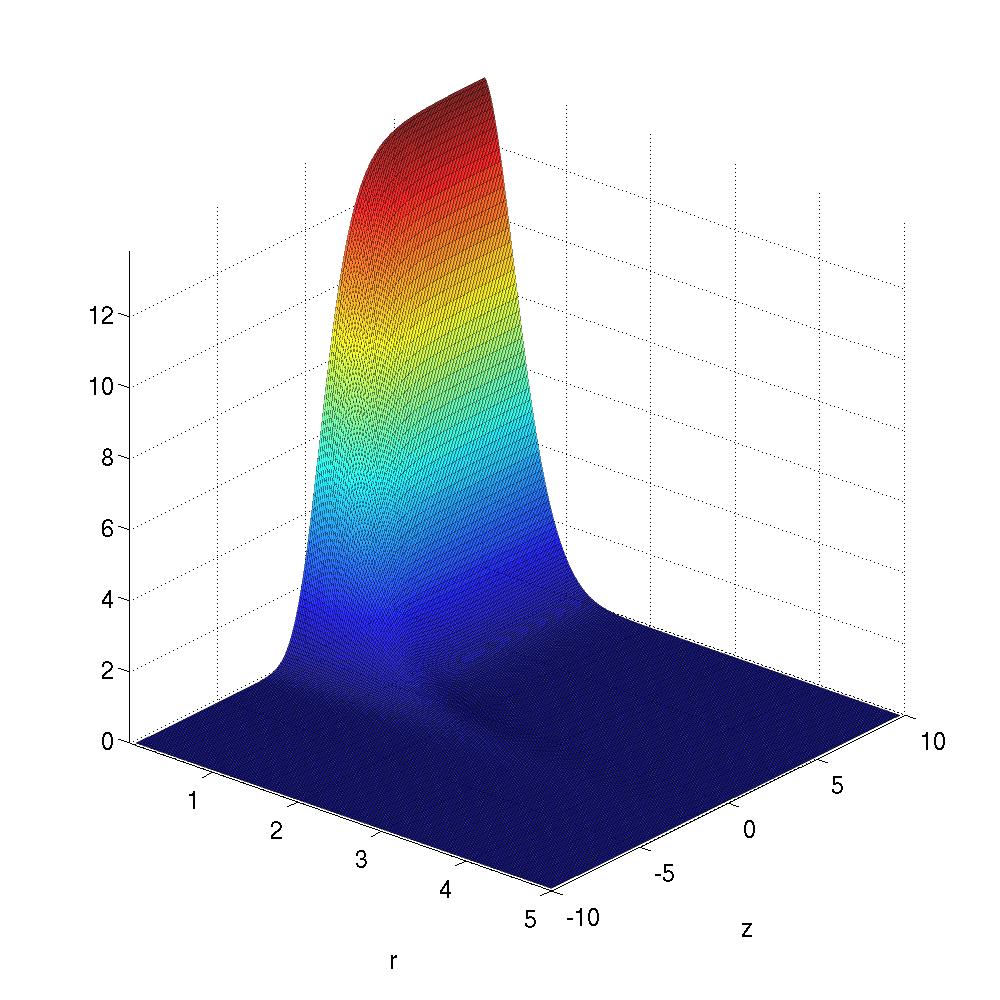}}}
\caption{${\rm O}(4)$ soliton junction of a domain wall (in the $z$
  direction) and a Skyrmion string carrying $\pi_3$ charge. The three
  panels show the field $n_4$, the energy density on a logarithmic
  scale and the charge density.}
\label{fig:o4}
\end{center}
\end{figure}

In Figs.~\ref{fig:o3} and \ref{fig:o4} are shown the numerical
solutions, the corresponding energy densities and charge densities,
for the O$(3)$ and O$(4)$ model, respectively. 

\begin{figure}[!thp]
\begin{center}
\mbox{
\subfloat{\includegraphics[width=0.49\linewidth]{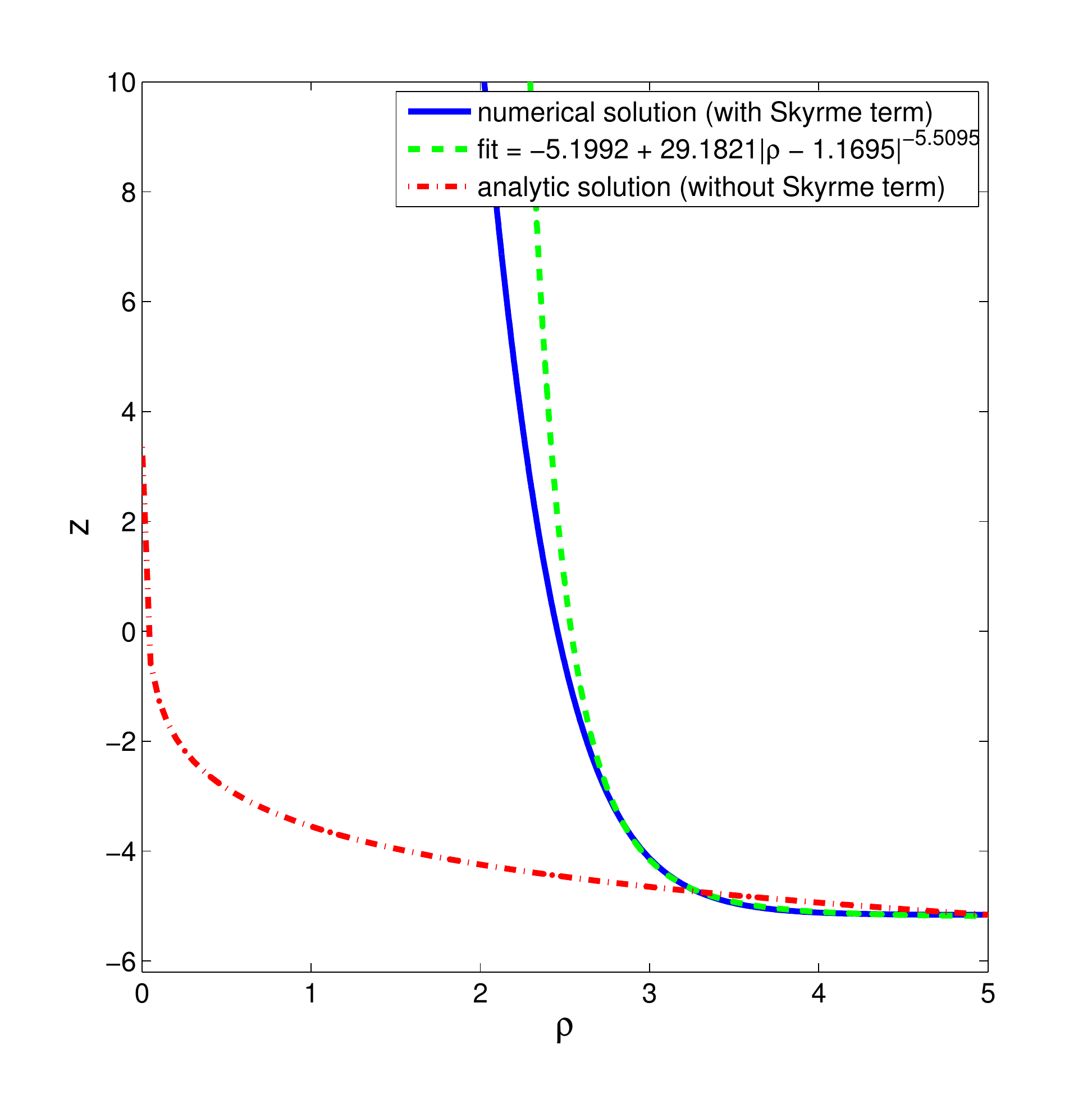}}
\subfloat{\includegraphics[width=0.49\linewidth]{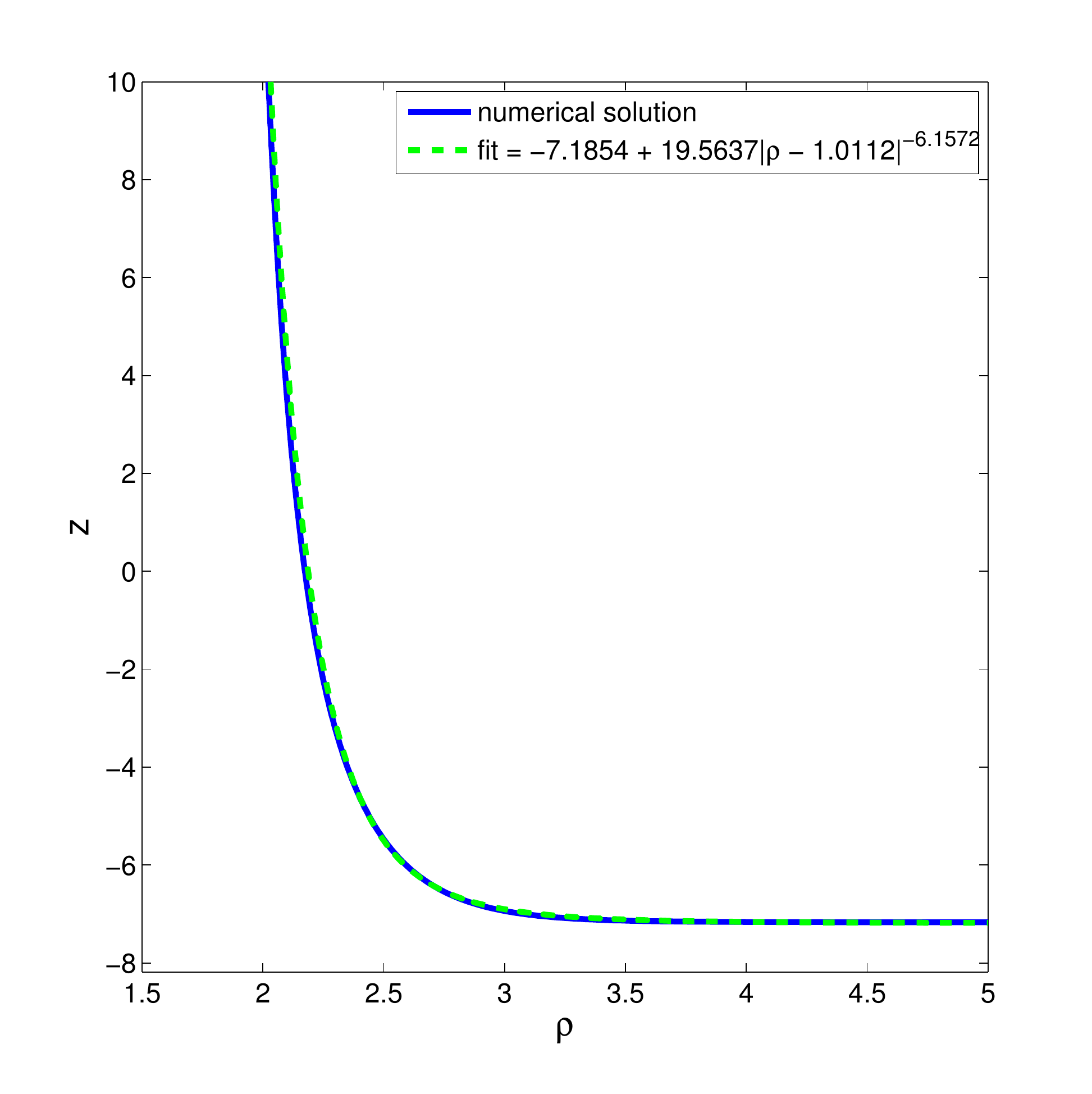}}}
\caption{Contour line of the soliton junction ($\cos f=0$), describing
  the bending of the domain wall due to the attached string. The
  left-hand (right-hand) side panel is for the O$(3)$-case
  (O$(4)$-case) and the fit is made with the numerical data in the
  region $z<-3$ ($z<-4$). }  
\label{fig:wallbending_skyrmemodel}
\end{center}
\end{figure}

Interestingly, the shape of the wall junction is altered somewhat
drastically. In Fig.~\ref{fig:wallbending_skyrmemodel} is shown the
contour line of the field $n_N=0$ in the $(\rho,z)$-plane. In the O(3)
case, a comparison with the analytic sigma-model solution is shown
with the red dashed-dotted line. A fit of the asymptotic part of the
junction is also shown with a green dashed line. The function is found
to be a power function, 
\beq
{\rm contour} = z_0 + b |\rho - w|^p,
\label{eq:fitfunction}
\eeq
where $z_0$ is the position of the domain wall, $b$ is a
proportionality constant, $w$ is the width of the string and $p$ is
the sought-after power describing the bending of the domain wall. The
fits find $p$ to be about 5-6.

\section{The 6th order model}\label{sec:6thmodel}

In this section, we want to consider $N=5,6$ which requires at least a
sixth-order derivative term, in order for the string to have a finite
thickness, see Eq.~\eqref{eq:scaling_argument}. 

\begin{figure}[!thp]
\begin{center}
\captionsetup[subfloat]{labelformat=empty}
\mbox{
\subfloat[$n_5=\cos f$]{\includegraphics[width=0.33\linewidth]{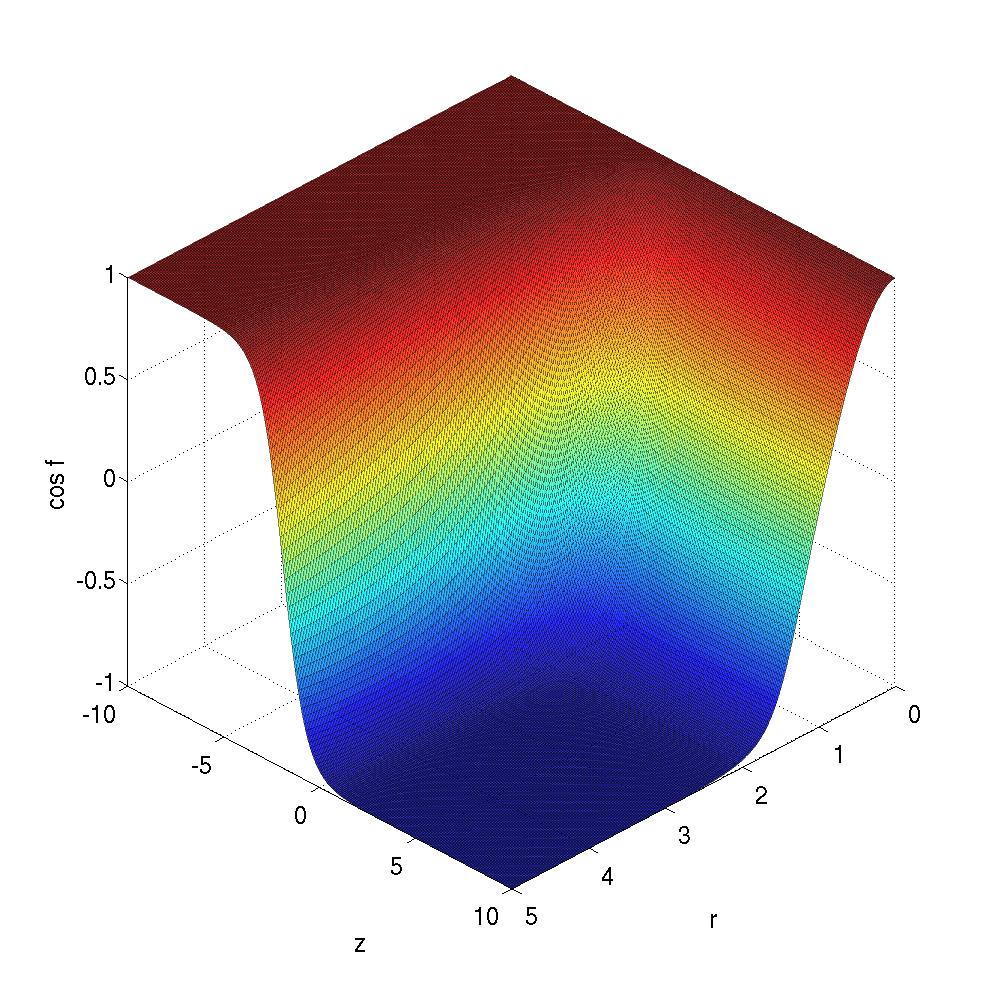}}
\subfloat[$\log(1+\mathcal{E})$]{\includegraphics[width=0.33\linewidth]{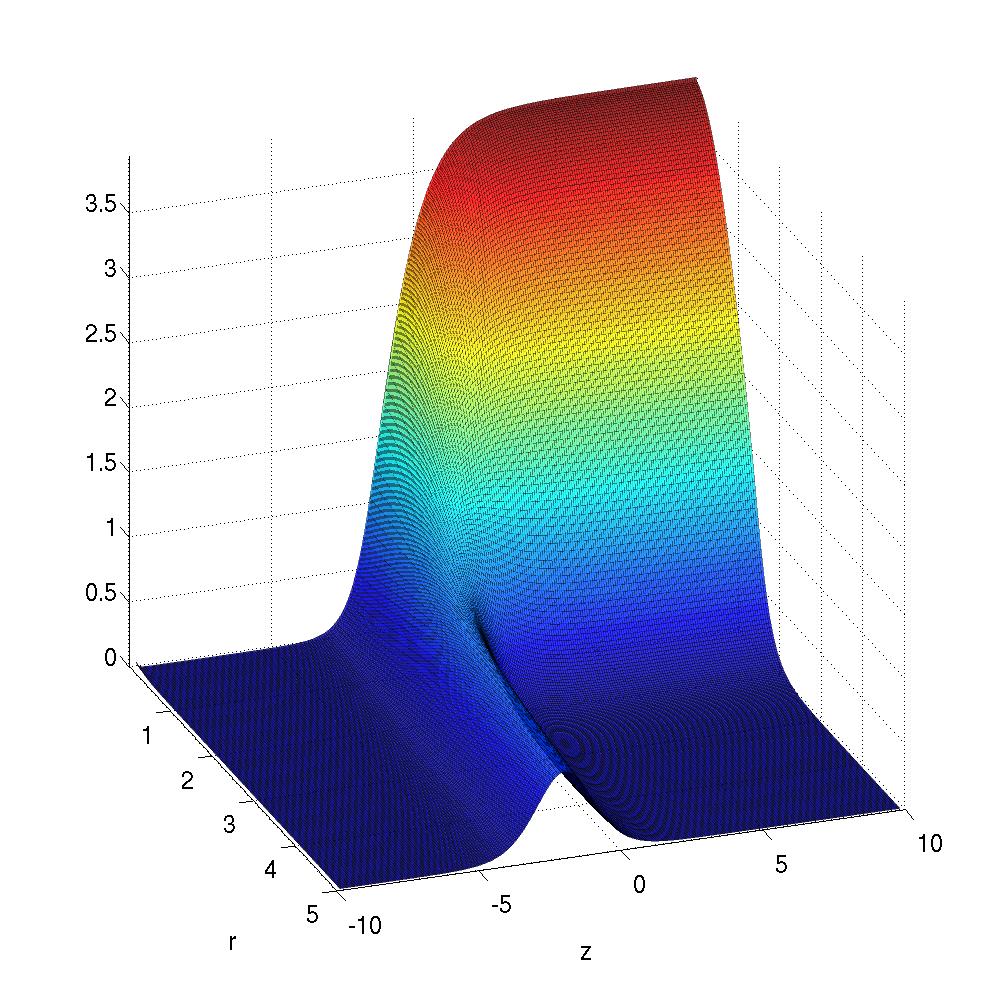}}
\subfloat[$\pi_4$ charge]{\includegraphics[width=0.33\linewidth]{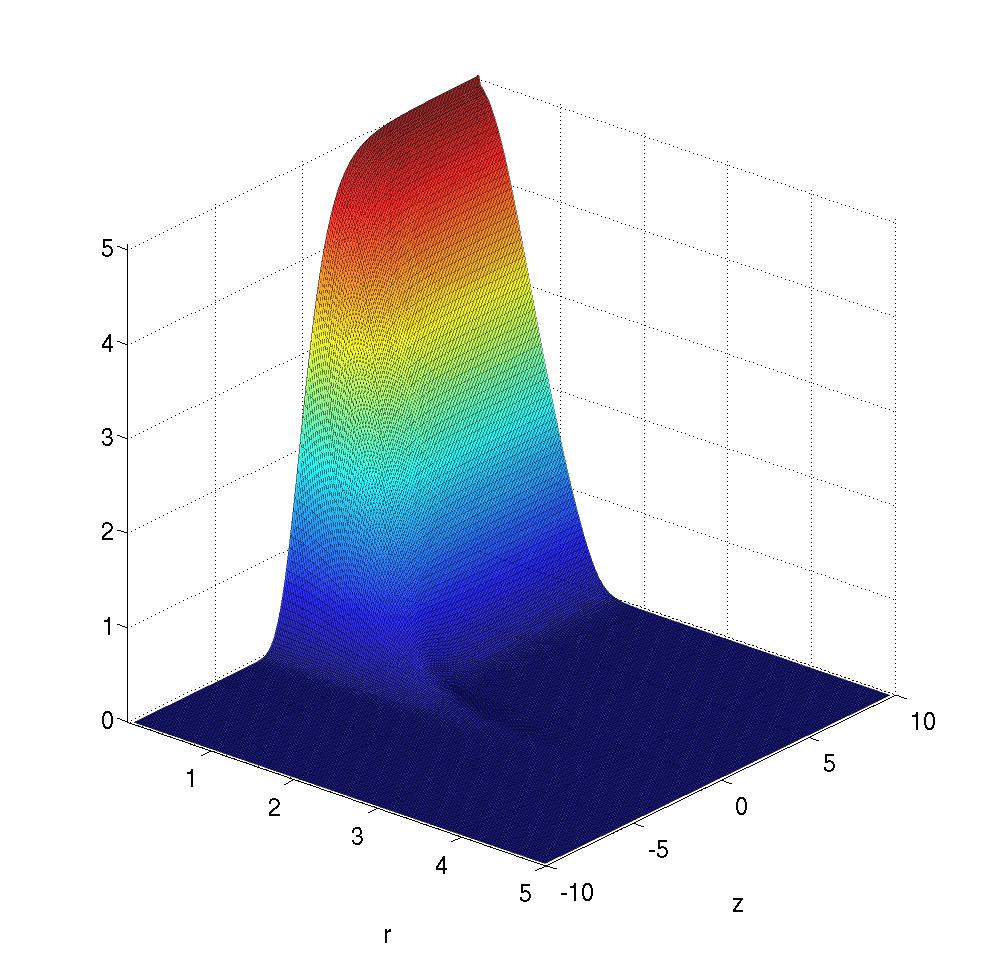}}}
\caption{${\rm O}(5)$ soliton junction of a domain wall (in the $z$
  direction) and a string carrying $\pi_4$ charge. The three
  panels show the field $n_5$, the energy density on a logarithmic
  scale and the charge density. }
\label{fig:o5}
\end{center}
\end{figure}

\begin{figure}[!thp]
\begin{center}
\captionsetup[subfloat]{labelformat=empty}
\mbox{
\subfloat[$n_6=\cos f$]{\includegraphics[width=0.33\linewidth]{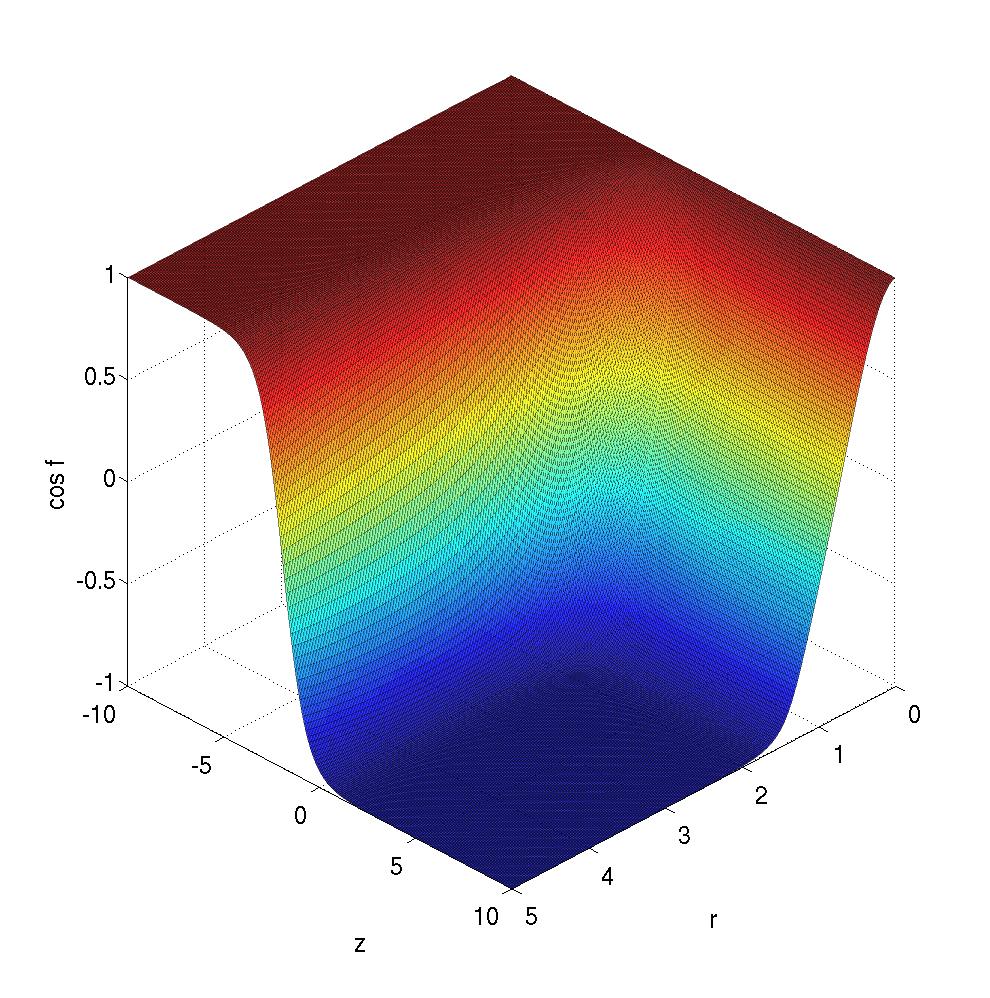}}
\subfloat[$\log(1+\mathcal{E})$]{\includegraphics[width=0.33\linewidth]{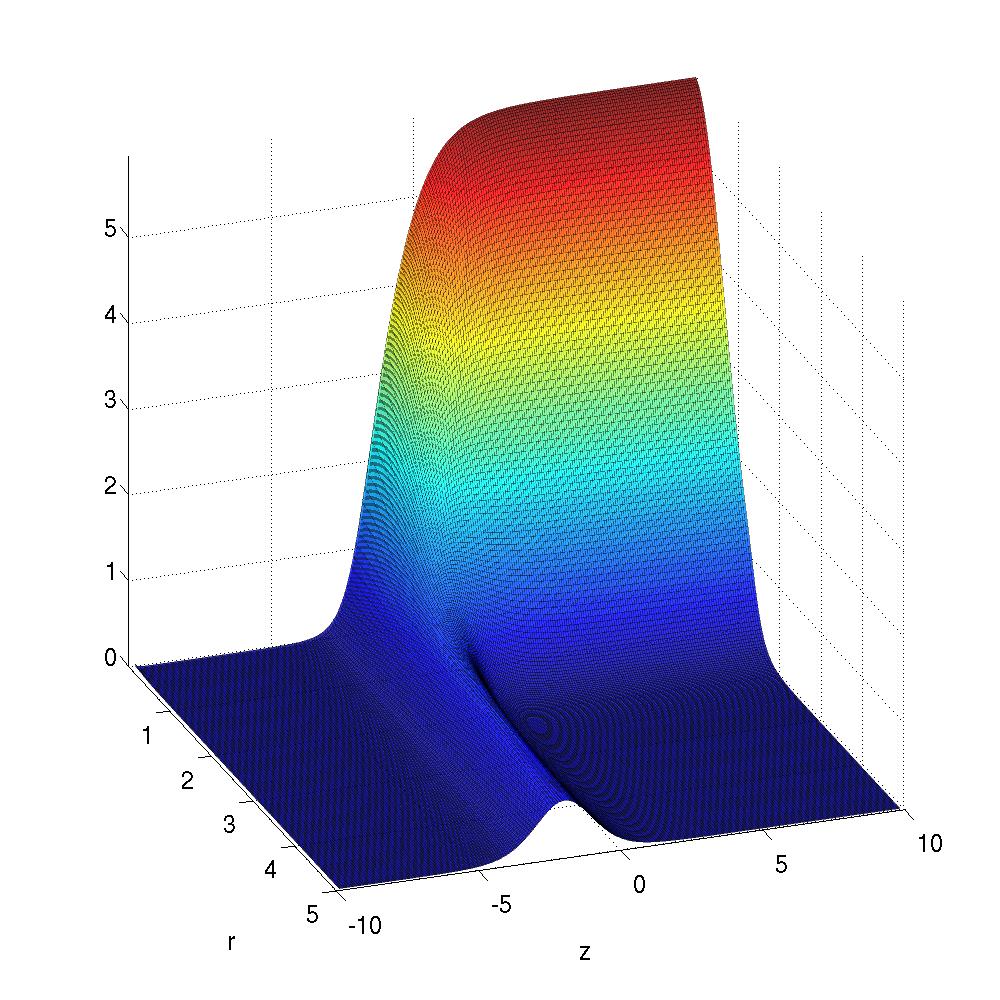}}
\subfloat[$\pi_5$ charge]{\includegraphics[width=0.33\linewidth]{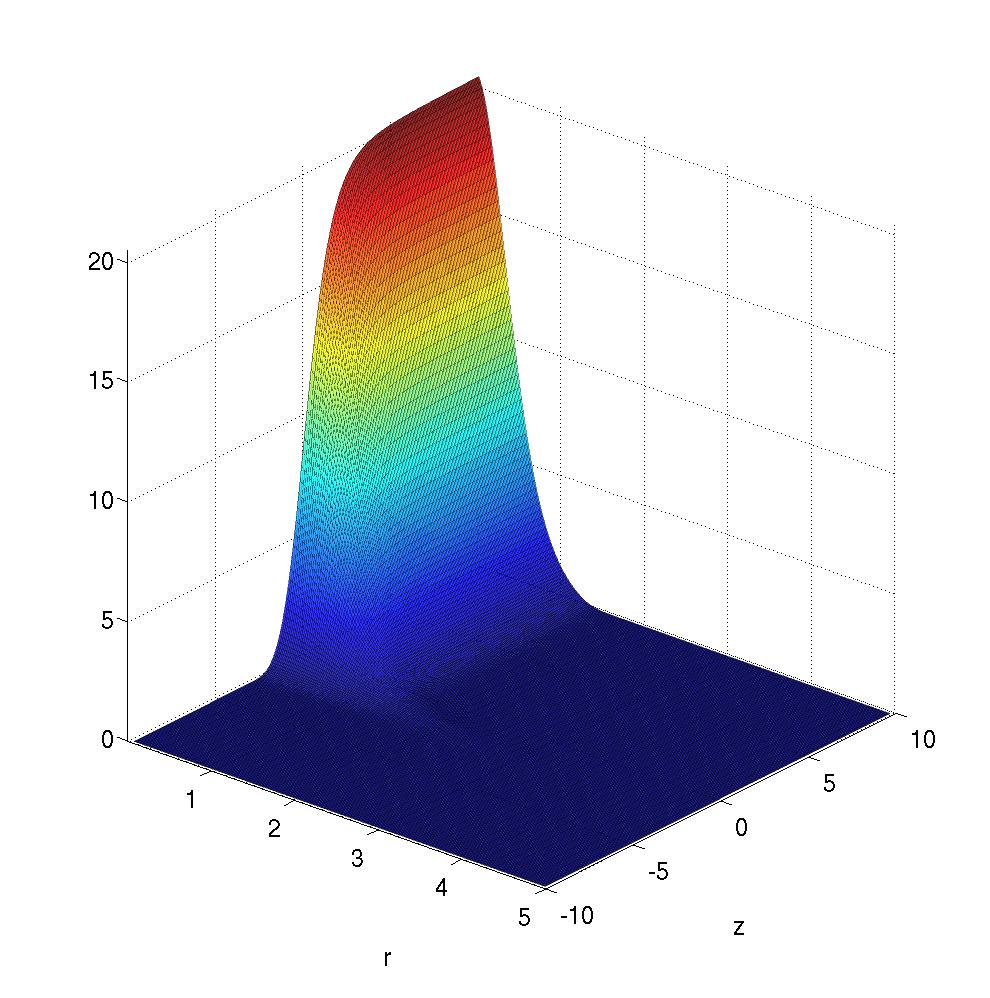}}}
\caption{${\rm O}(6)$ soliton junction of a domain wall (in the $z$
  direction) and a string carrying $\pi_5$ charge. The three
  panels show the field $n_6$, the energy density on a logarithmic
  scale and the charge density. } 
\label{fig:o6}
\end{center}
\end{figure}

We will again use a finite-difference scheme on a quadratic square
lattice with $256^2$ lattice sites and relax initial guesses with the
relaxation method.
In Figs.~\ref{fig:o5} and \ref{fig:o6} are shown the numerical
solutions, the corresponding energy densities and charge densities,
for the O$(5)$ and O$(6)$ model, respectively. 

\begin{figure}[!thp]
\begin{center}
\mbox{
\subfloat{\includegraphics[width=0.49\linewidth]{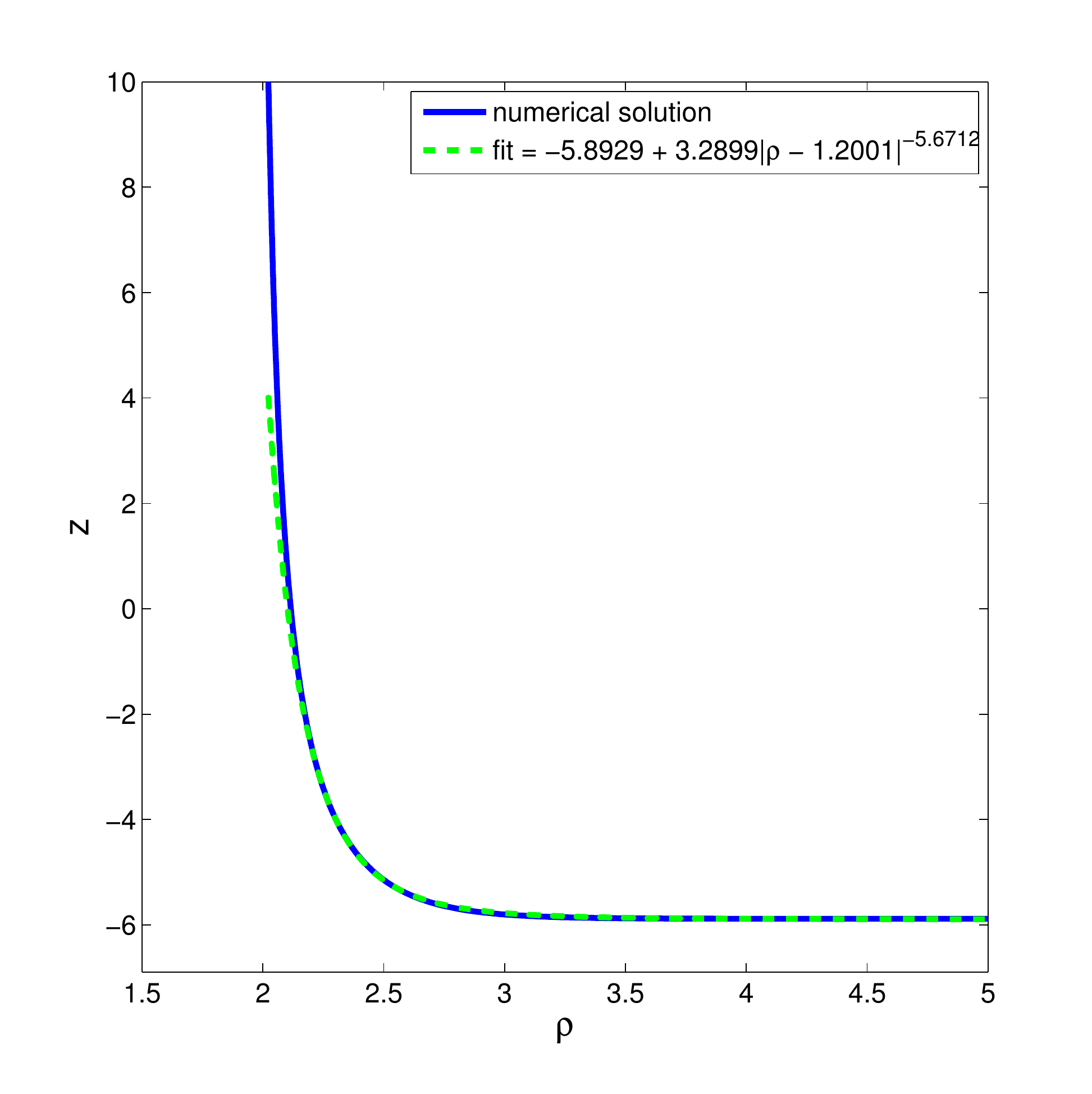}}
\subfloat{\includegraphics[width=0.49\linewidth]{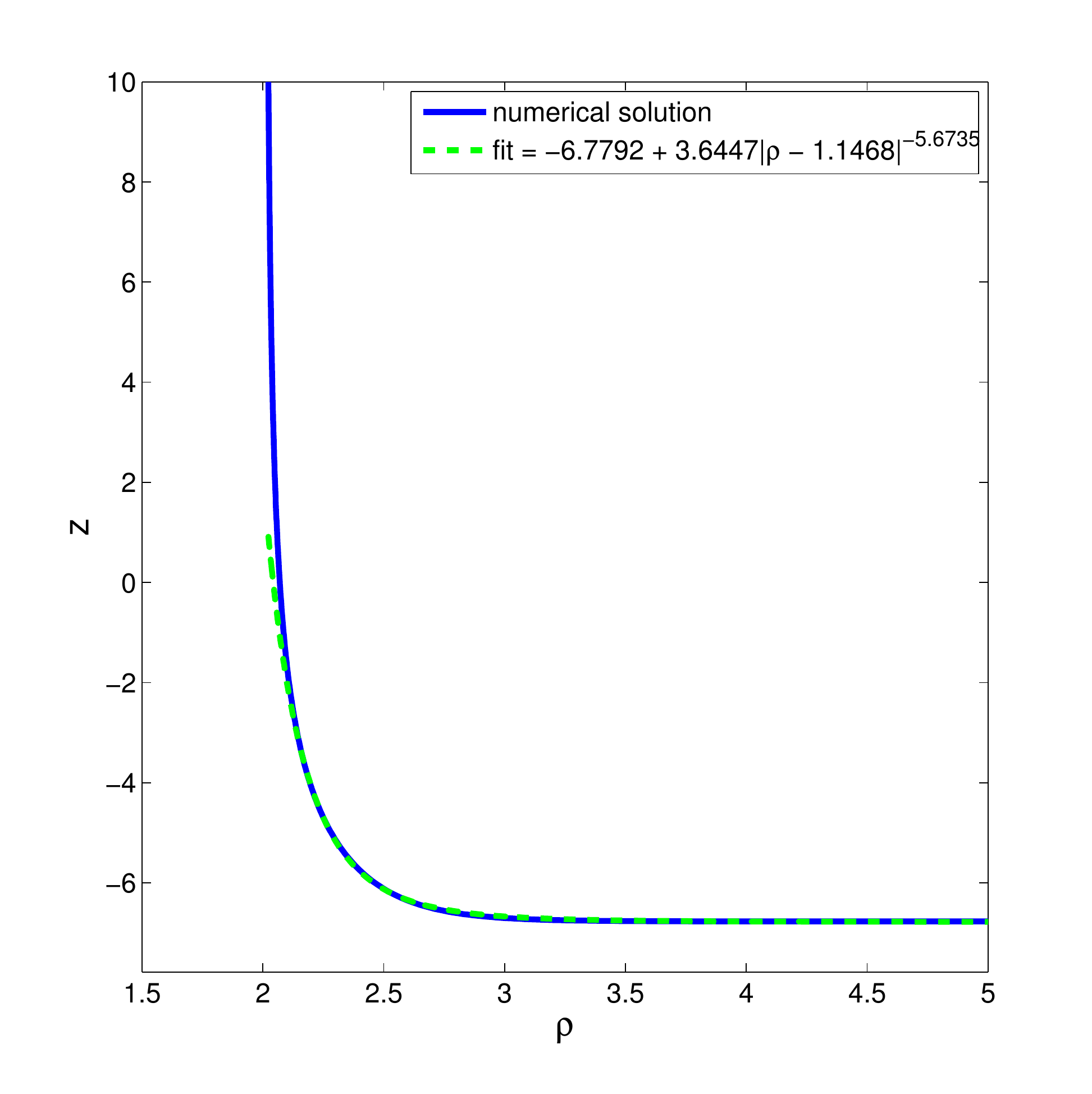}}}
\caption{Contour line of soliton junction ($\cos f=0$), describing the
  bending of the domain wall due to the attached string. The left-hand
  (right-hand) side panel is for the O$(5)$-case (O$(6)$-case) and the
  fit is made with the numerical data in the region $z<-3$. }
\label{fig:wallbending_6thmodel}
\end{center}
\end{figure}

We consider again the shape of the wall junction and show the contour
line of the field $n_N=0$ in the $(\rho,z)$-plane as well as a fit of
the type \eqref{eq:fitfunction} in
Fig.~\ref{fig:wallbending_6thmodel}.
The powers $p$ are again fitted to be about 5-6. 

\begin{figure}[!thp]
\begin{center}
\includegraphics[width=0.66\linewidth]{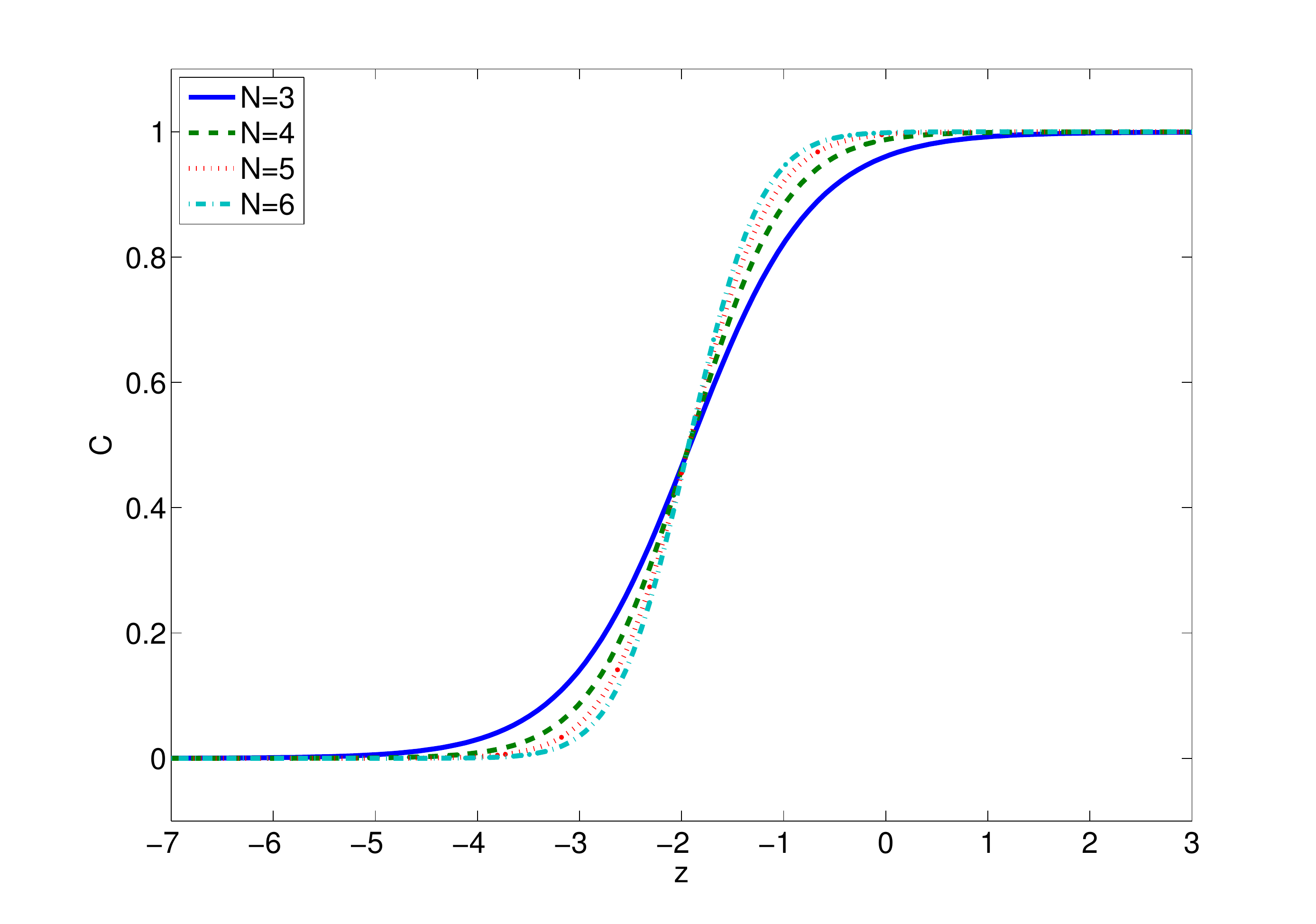}
\caption{String charge $C$ as function of $z$. Far away from the
  domain wall it is 0 and 1, on the left and right-hand
  side of the domain wall, respectively. }
\label{fig:stringcharge}
\end{center}
\end{figure}

Finally, we consider the string charge which as function of $z$ has to
interpolate from a full charge (1) to zero across the wall
junction. Hence we plot Eq.~\eqref{eq:stringcharge} across the
domain-wall junction for all the obtained solutions in 
Fig.~\ref{fig:stringcharge}. It is seen that the transition becomes
more steep with increasing $N$, which may be expected just on
dimensional grounds. 

\section{Summary and Discussion} \label{sec:summary}

We have constructed D-brane solitons, 
composite solitons of strings ending on a domain wall
in an O($N$) model with a higher-derivative term 
in $d=N+1$ dimensions. 
For $N=3$, it is a baby-Skyrmion string 
ending on a domain wall in $d=3+1$, 
while for $N=4$, it is a Skyrmion string ending on a domain wall 
in $d=4+1$.
In general, 
a string supported by the $\pi_{N-1}$ topological charge 
ends on a domain wall and bends the domain wall like $1/\rho^{\#}$. 
In this paper, we have considered only $N=3,4,5,6$ where we have
calculated the needed terms explicitly. A generalization to higher $N$
can be carried out by considering the higher-order derivative term
\beq
\mathcal{L}_{2m}=-\left(\p_{\mu_1}^{[a_1}\cdots\p_{\mu_m}^{a_m]}\right)^2,
\eeq
where $\mu_i=1,\ldots,N$; $a_i=1,\ldots,N$; $i=1,\ldots,m$ and 
$m=\lceil N/2\rceil$ ($\lceil x\rceil$ rounds up a real number to the
nearest integer).

In this paper, 
strings are of various codimensions depending on the dimension,
while ``D-branes'' are all of domain-wall type, that is, of
codimension one. 
For instance, a vortex string (of codimension two) 
with a confined Skyrmion was constructed in
Refs.~\cite{Gudnason:2014hsa,Gudnason:2014jga}. 
The generalization to higher codimensions for D-branes 
remains a future problem. 
In supersymmetric theories, 
all possible composite BPS solitons 
were classified in Ref.~\cite{Eto:2005sw},
which may be useful for this study.

Field theory D-branes beyond the semi-classical approach were studied
in Ref.~\cite{Tong:2005nf}, 
in which the bulk-boundary correspondence was proposed. 
Our study could be applied to that direction as well.

\section*{Acknowledgments}

The work of MN is supported in part by Grant-in-Aid for Scientific Research 
(No.~25400268) 
and by the ``Topological Quantum Phenomena'' 
Grant-in-Aid for Scientific Research 
on Innovative Areas (No.~25103720)  
from the Ministry of Education, Culture, Sports, Science and Technology 
(MEXT) of Japan. 
S.~B.~G.~thanks Keio University for hospitality during which this
project took shape. S.~B.~G.~also thanks Institute of Modern Physics
(CAS), Lanzhou, China for hospitality. 


\end{document}